\documentclass[sigconf,authorversion]{acmart} 

\AtBeginDocument{%
  }

\copyrightyear{2023}
\acmYear{2023}
\setcopyright{licensedusgovmixed}\acmConference[CHI '23]{Proceedings of the 2023 CHI Conference on Human Factors in Computing Systems}{April 23--28, 2023}{Hamburg, Germany}
\acmBooktitle{Proceedings of the 2023 CHI Conference on Human Factors in Computing Systems (CHI '23), April 23--28, 2023, Hamburg, Germany}
\acmPrice{15.00}
\acmDOI{10.1145/3544548.3580869}
\acmISBN{978-1-4503-9421-5/23/04}

\acmSubmissionID{2445}

\begin{document}

\title{\bf \texttt{MetaExplorer}: Facilitating Reasoning with Epistemic Uncertainty in Meta-analysis}

\author{Alex Kale}
\email{kalea@uchicago.edu}
\affiliation{%
  \institution{University of Chicago}
  \city{Chicago}
  \state{Illinois}
  \country{USA}
}

\author{Sarah Lee}
\email{sarahlee@stottlerhenke.com}
\affiliation{%
  \institution{Stottler Henke Assoc.}
  \city{Seattle}
  \state{Washington}
  \country{USA}
}

\author{Terrance Goan}
\email{goan@stottlerhenke.com}
\affiliation{%
   \institution{Stottler Henke Assoc.}
  \city{Seattle}
  \state{Washington}
  \country{USA}
}

\author{Elizabeth Tipton}
\email{tipton@northwestern.edu}
\affiliation{%
 \institution{Northwestern University}
  \city{Evanston}
  \state{Illinois}
  \country{USA}
}

\author{Jessica Hullman}
\email{jhullman@northwestern.edu}
\affiliation{%
  \institution{Northwestern University}
  \city{Evanston}
  \state{Illinois}
  \country{USA}
}

\renewcommand{\shortauthors}{Kale et al.}

\begin{abstract}
  Scientists often use meta-analysis to characterize the impact of an intervention on some outcome of interest across a body of literature.
  However, threats to the utility and validity of meta-analytic estimates arise when scientists average over potentially important variations in context like different research designs.
  Uncertainty about quality and commensurability of evidence casts doubt on results from meta-analysis, yet existing software tools for meta-analysis do not necessarily emphasize addressing these concerns in their workflows. 
  We present \texttt{MetaExplorer}, a prototype system for meta-analysis that we developed using iterative design with meta-analysis experts to provide a guided process for eliciting assessments of uncertainty and reasoning about how to incorporate them during statistical inference.
  Our qualitative evaluation of \texttt{MetaExplorer} with experienced meta-analysts shows that imposing a structured workflow both elevates the perceived importance of epistemic concerns and presents opportunities for tools to engage users in dialogue around goals and standards for evidence aggregation.
\end{abstract}



\begin{teaserfigure}
    \includegraphics[width=\textwidth]{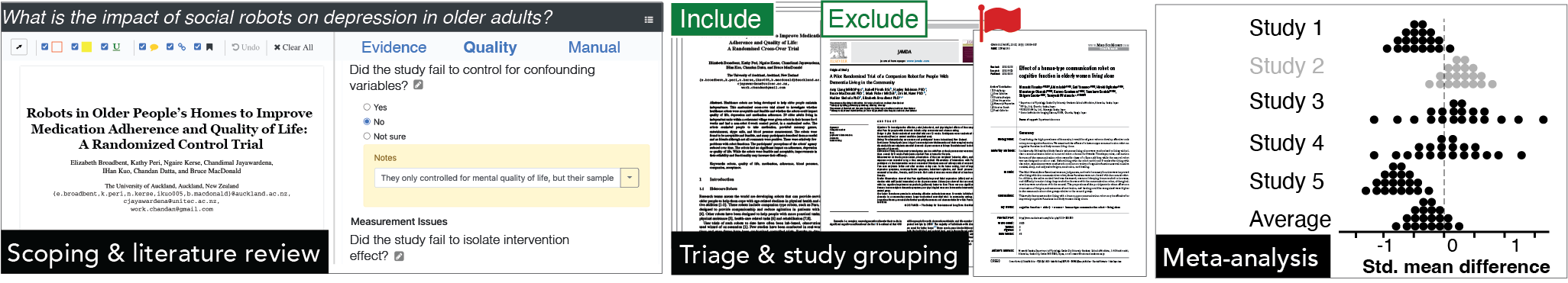}
    \caption{\texttt{MetaExplorer} provides a guided process for literature review and meta-analysis with an emphasis on documenting sources of epistemic uncertainty and choosing how to address them during statistical inference. The workflow proceeds in stages from \textbf{Scoping and literature review} where \texttt{MetaExplorer} elicits information about each study, to \textbf{Triage and study grouping} where the user resolves sources of epistemic uncertainty, and finally to \textbf{Meta-analysis} where the user views results alongside contextualizing uncertainty.
    }
    \label{fig:teaser}
    \Description{An overview of MetaExplorer’s workflow: scoping and literature review; triaging epistemic uncertainties by including, excluding, flagging, or grouping study results; meta-analyzing groups of studies to produce average effect sizes summarizing a corpus of literature.}
\end{teaserfigure}

\maketitle

\section{Introduction}
Summarizing a corpus of scientific literature poses challenges, even for seasoned researchers with domain knowledge.
This is especially difficult when the purpose of the review is to be both systematic---including all relevant studies, not just familiar ones---and to inform a decision or practice. 
The process of extracting and combining data from multiple studies is referred to as \textit{meta-analysis}, and involves searching for and finding relevant papers, ensuring they answer the research question, extracting information (including statistical estimates) from each paper, and combining this information into summary measures.  Meta-analyses are common in a variety of fields, including medicine, social welfare, economics, and education, where results from such reviews are perceived as central to ``evidence-based'' decision making.

To understand why such reviews are difficult, \textit{imagine a scenario} where a human-computer interaction researcher, Kara, consults for a retirement community about whether purchasing social robots would improve the psychological wellbeing of residents.
Seeing demonstrations of early social robots (e.g.,~\cite{Moyle2013}) excited Kara's clients, but before purchasing anything, they want Kara to verify whether empirical research supports the idea that social robots can improve mental health indicators such as depression. 
Since social robots are a relatively new invention, Kara anticipates that there are not many controlled studies on them yet, but sets out to review this small literature.
She first searches for relevant papers that include ``social robots'' and ``depression,'' then screens such studies to ensure they answer her research questions.
These papers might report measures of depression for groups that use social robots compared to those that do not, or compare depression within individuals before versus after using social robots. 
Some might report positive effects and others negative effects, with effect sizes varying from negligible to moderate. 
Kara could use \textit{meta-analysis} to summarize the evidence as an average effect and its variation across studies. 
However, to do so, Kara must make difficult judgments about the quality of individual study results, whether different measures can be meaningfully combined, and whether evidence from a given study will generalize to the target context one wants to make an inference about. 

Known pitfalls when scientists use meta-analysis to estimate intervention effects and inform decisions (e.g., about purchasing social robots) include focusing too strongly on the average effect, and interpreting this average as a `true' fixed and universal effect~\cite{bryan2021behavioural,szaszi2022no}, despite many systematic reviews (and common-sense expectations about effects in the world~\cite{bryan2021behavioural}) 
suggesting a range of effects that vary across contexts and study designs. 
Conceptual frameworks can help by breaking down these variations by sources, such as Methods used, Units studied, Treatment versions, Outcomes measured, and Settings considered (MUTOS)~\cite{becker2017improving,Manski2019-MA}, differentiating, for example, issues of internal 
and external validity~\cite{Shadish2002-mutos}.
Scientists must weigh these concerns and consider which studies should be included or grouped together in their meta-analysis, and may even decide that meta-analysis is not appropriate for their corpus.

In particular, scientists routinely struggle to account for the impact of \textit{epistemic uncertainty} on results in evidence aggregation~\cite{greenland2001quality,higgins2002heterogeneity,turner2009bias}.
Unfortunately, current software tools for meta-analysis can limit scientists' ability to externalize concerns about these uncertainties in ways that clearly inform statistical inference~\cite{kale2019decision}. 
This makes scientific review and meta-analysis a useful application area for investigating how software can represent and facilitate reasoning about epistemic uncertainty in evidence aggregation more broadly.

We contribute \texttt{MetaExplorer}, a prototype system providing a workflow for eliciting sources of epistemic uncertainty from scientists during literature review and helping them respond to these during meta-analysis. 
\texttt{MetaExplorer} combines several features uncommon among meta-analysis tools: 
(1) a guided triage process for reasoning about how epistemic uncertainties may threaten inferential validity; (2) an exploratory visualization for reasoning about quantified inferential uncertainty alongside unquantified uncertainty; and (3) visualizations of inferential uncertainty framed as possible outcomes.
We created \texttt{MetaExplorer} using an iterative user-centered design process with frequent feedback from experienced (n = 5) and expert (n = 3) meta-analysts, and further evaluate it in guided-tour interviews with participants who are experienced and knowledgeable about systematic review and meta-analysis (n = 12).
These participants reflect the population of intended users for \texttt{MetaExplorer}, scientists who conduct meta-analysis to answer practical questions.
Our analysis of these interviews reveals that guided-process tools like \texttt{MetaExplorer} seem to derive their benefits and drawbacks in part from challenging users' conceptualizations of analysis tasks.
Elevating concerns about epistemic uncertainty, from optional to focal for meta-analysis, requires a structured procedure for handling them. However, in order to flexibly accommodate standards in a variety of scientific domains, these procedures must also adapt in dialogue with the user's analysis goals.
Our findings about \texttt{MetaExplorer} point toward new ways of designing \textit{analysis tools as partners in deliberation} about ambiguity in data analysis. 
\section{Background}
We contextualize our work on \texttt{MetaExplorer} in relation to other tools supporting meta-analysis and literature review, interdisciplinary perspectives on reasoning with epistemic uncertainty, and visualization techniques for showing quantified inferential uncertainty.

\subsection{Supporting meta-analysis \& scientific review}
Constructing a meta-analysis from a scientific review entails judging what demarcates populations of studies: meta-analysis assumes that study results are sampled from a statistical population of studies and thus can be meaningfully averaged~\cite{cooper2019handbook,lipsey2001practical}.
A sampled corpus of literature may contain substantial heterogeneity, which is typically accounted for by using a hierarchical (a.k.a. random effects) model to separate variance between and within studies, representing heterogeneity and residual inferential uncertainty, respectively~\cite{cooper2019handbook}.
\texttt{MetaExplorer} employs hierarchical models as they are a ``gold standard''~\cite{lipsey2001practical}, and emphasizes reasoning about epistemic uncertainty in literature review in terms of what results to aggregate.

Most software tools for scientific review do not support meta-analysis but are flexible in enabling users to document epistemic uncertainty.
Tools such as EPPI-Reviewer~\cite{eppi-reviewer}, Distiller SR~\cite{distillersr}, Covidence~\cite{covidence}, and Rayyan~\cite{rayyan} provide \textit{document analysis interfaces} focused on annotating and coding articles.
Such tools include reference managers like Mendeley~\cite{mendeley} and Zotaro~\cite{zotero}.
Some document analysis tools focus specifically on helping researchers evaluate quality of evidence---e.g., Newcastle-Ottawa scale~\cite{Newcastle-Ottawa-Scale}, GRADE criteria~\cite{Grade}, Cochrane Risk of Bias Assessment~\cite{Cochrane-riskofbias}, or Jadad scale~\cite{Jadad-scale}. 
Many of these are structured like checklists, but focus on disjoint or partially overlapping sets of epistemic issues (e.g., risk of bias vs generalizability) and produce different output formats, making it challenging to collate results across scales. 
Such quality assessment scales are sometimes integrated as options in document annotation tools---e.g., EPPI-Reviewer~\cite{eppi-reviewer} includes Cochrane Risk of Bias Assessment, and other tools enable such assessments if the user manually implements them~\cite{distillersr,rayyan}. 
Extending these designs, \texttt{MetaExplorer}'s evidence extraction tool interleaves questions typically required for scientific review and meta-analysis with a new quality assessment form.
\texttt{MetaExplorer}'s quality assessment synthesizes existing scales to cover a union of sources of epistemic uncertainty addressed by other quality assessments and produces a unified output format, where every question receives an answer of `yes' (there is an issue), `no', or `not sure' accompanied by notes documenting the user's rationale. 
The new scale and unified output format streamline quality assessment during literature review and resolve difficulties around collating results from existing scales.

In current practice, scientists often need to context-switch to \textit{statistical tools} to perform a meta-analysis.
For example, CMA~\cite{cma} and MIX~\cite{mix} are implemented as add-ons for Microsoft Excel~\cite{msexcel}.
These tools also include R packages such as metafor~\cite{metafor} and meta~\cite{meta}.
Others have built cross-platform meta-analysis software such as MetaWin~\cite{metawin}, OpenMetaAnalyst~\cite{OpenMetaAnalyst}, Annotation Graphs~\cite{Zhou2017}, and MetaInsight~\cite{Owen2019}, which are similarly focused on estimating meta-analytic averages.
To our knowledge, Cochrane's RevMan~\cite{revman} is unique in supporting both meta-analysis and document analysis, however, Cochrane's Risk of Bias Assessment~\cite{Cochrane-riskofbias} is optional and focuses on internal validity only.
\texttt{MetaExplorer} explores how to help scientists transition from literature review to meta-analysis in one tool, with an emphasis on quality assessment.

\subsection{Reasoning with epistemic uncertainty in data analysis}
Epistemic uncertainty is prevalent in data analysis decisions---e.g., how to model a dataset.
The necessity of such judgments and problem of how they impact the results of analysis has been dubbed ``researcher degrees of freedom''~\cite{Wicherts2016}.
Pre-registration~\cite{Nosek2018-prereg} and multiverse analysis~\cite{Steegen2016,Simonsohn2015} aim to guard against threats to validity by making analysis choices explicit.
However, supporting such procedures requires software representation of epistemic uncertainty around analysis choices~\cite{kale2019decision,Liu2020-analysis-decisions}.
Recent work in human-computer interaction addresses this challenge primarily by attempting to guide analysts in selecting among possible models~\cite{Jun2022-tisane,Liu2021-boba,wacharamanotham2015a,Zhou2017} and surfacing provenance about measurements~\cite{McCurdy2018_infovis_ie-framework,Silva2011-vistrails}.

One major problem in designing software to help scientists reason about epistemic uncertainty that threatens meta-analysis is that scientists conducting research synthesis tend document these sources of uncertainty (e.g., study quality concerns) in ad hoc ways such as by writing them in lab notebooks~\cite{kale2019decision}.
Subsequently, they struggle to integrate these uncertainties into their statistical inferences through practices such as sensitivity analysis~\cite{kale2019decision,Liu2020-analysis-decisions}, perhaps because software tools are not designed to help to maintain awareness of uncertainty~\cite{Sacha2016}.
Another major challenge for scientists is deciding how to respond to epistemic uncertainty.
Prior work~\cite{boukhelifa2017,kale2019decision} characterizes strategies for resolving uncertainty in terms of ``suppressing'' or ignoring it versus ``reducing'' or incorporating it into analysis through mechanisms such as statistical modeling.
For example, in meta-analysis, if a study result may be biased, scientists should check the impact on results when removing it from their model (i.e., sensitivity analysis) to see if their statistical inference is robust to potential study quality issues.
Studies that seem to measure different constructs should be modeled separately for clarity of interpretation.
Study results that are not applicable to the target context a scientist wants to make an inference about may still be informative, but including them in meta-analysis leads to estimates that may not generalize.
\texttt{MetaExplorer} extends prior work on representing and managing epistemic uncertainty by structuring the meta-analysis workflow around resolving ambiguities of interpretation that influence decisions in evidence aggregation.

\subsection{Visualizing inferential uncertainty}
Conventional techniques for uncertainty visualization require an approach to uncertainty quantification that produces distributions or boundaries to show, and thus they do not address unquantified epistemic uncertainties (see Section 2.2).
For example, among the most common applications of uncertainty visualization are confidence intervals showing \textit{inferential uncertainty} about estimates from a statistical model~\cite{Cumming2014,Manski2018b,taylor_guidelines_1994}, such as those in forest plots generated by most meta-analysis software (e.g.,~\cite{cma,mix,metafor,meta,metawin,OpenMetaAnalyst,revman,Owen2019}).
Despite their prevalence, previous work on statistical cognition~\cite{Belia2005,Soyer2012,Hoekstra2014} and uncertainty visualization~\cite{Correll2014,Hullman2015,Kale2019-hops,Kay2016,Fernandes2018,Kale2021strategies,Castro2022} finds widespread misinterpretation of interval representations of uncertainty.
Drawing on cognitive science suggesting benefits of framing probabilities as frequencies of events~\cite{Gigerenzer1995,Hoffrage1998}, alternative techniques, such as \textit{quantile dotplots}~\cite{Kay2016} show percentiles of the underlying univariate distribution as stacked dots, enabling users to reason about probabilities by counting.
Multiple empirical evaluations to date find that quantile dotplots support visual statistical inferences better than interval representations of uncertainty~\cite{Kay2016,Fernandes2018,Kale2021strategies}.
Following previous work, we adopt quantile dotplots as an alternative to confidence intervals in \texttt{MetaExplorer}'s interactive forest plot (see Section 4.1.5).
\section{Designing for meta-analysis}
Our aim in prototyping \texttt{MetaExplorer} was a guided process for meta-analysis that elicits sources of epistemic uncertainty alongside effect size statistics during literature review and propagates these uncertainties, making it easier to conduct meta-analysis with epistemic uncertainty as a primary consideration. Our primary design goals are: 
\begin{itemize}
    \item \textbf{Make epistemic uncertainties explicit.} A tool should elicit and explicitly represent sources of epistemic uncertainty about scientific literature.
    \item \textbf{Non-optional quality assessment.} A tool should integrate quality assessment with meta-analysis as a non-optional procedure, without extending the duration of scientific review.
    \item \textbf{Propagating concerns about study quality.} A tool should collate unquantified epistemic uncertainty in ways that can inform statistical modeling, without overwhelming the user.
    \item \textbf{Sensitivity analysis.} A tool should support exploration of possible inferences a user could reasonably make in a meta-analysis (e.g., by including different sets of results).
\end{itemize}

\subsection{Design process}
We arrived at the above design guidelines for \texttt{MetaExplorer} through an iterative user-centered design process.
We frequently gathered feedback from \textit{potential users}, initially running think-aloud pilot interviews with a paper prototype of the evidence extraction form to investigate what \textit{experienced meta-analysts} might want from a guided processes, and later eliciting informal feedback from \textit{experts in meta-analysis}. 
The distinction between experienced and expert meta-analysts reflects groups of participants containing some graduate students versus groups strictly comprised of PhDs with decades of experience specialized in research synthesis.
We sought input from both experienced meta-analysts and meta-analysis experts because we envisioned \texttt{MetaExplorer} as a tool to help scientists conduct quick meta-analyses to answer practical questions, and we wanted to support existing workflows and resolve pain points that scientists see as threats to the validity or feasibility of meta-analysis.
Feedback during this process led us to focus on how meta-analysis software can support epistemic uncertainty, rather than a broader set of meta-analysis considerations (e.g., dual review/collaboration features, support for more study designs). 

\textbf{Paper prototype sessions.} 
We created a paper prototype to elicit feedback on what questions belong in an interface for reviewing scientific articles, extracting effect size information, and eliciting epistemic uncertainty. 
We created the initial paper prototype drawing on best practices for meta-analysis and organizing principles for scientific review~\cite{lipsey2001practical,higgins2011cochrane}.
We instructed participants to think aloud while reviewing an article for inclusion in a hypothetical meta-analysis with the prototype.
In the second half of these interviews, we prompted an open-ended discussion with participants:
\begin{quote}
How can we generalize the process of evidence extraction through a form like the one you just used? What changes would you make to these materials? Are there things you consider when extracting evidence from articles which are not represented in the form? What sort of interface would be ideal for this task?
\end{quote}
This protocol first placed the participant in a ``work-like situation''~\cite{bodker1991cooperative} which allowed us to observe ``reflection in action''~\cite{schon1987educating}, enabling us to clarify what makes evidence extraction cumbersome. We conducted pilot interviews with five participants, who were scientists with previous experience with meta-analysis, recruited from our professional network.

\textbf{Informal feedback from experts.}
Throughout the development of \texttt{MetaExplorer} we met with experts in meta-analysis to share intermediate versions of the tool, so these experts could suggest changes to the tool and raise potentially challenging edge cases.
Our general process was to consider new features in a planning document, make a first-pass implementation, and test the interface by coding example articles.
This resulted in working examples of complete meta-analyses that we could use to demonstrate the tool for expert meta-analysts. 
We queried three experts from our professional network, who were all PhDs with extensive experience in evidence synthesis. 

\textbf{Summary of feedback.}
Feedback from pilot sessions and informal discussions with experts led to improvements in question wording and revealed issues that were especially challenging to reason about.
For example, pilot participants struggled to identify which of the many numbers reported in a paper were required for their meta-analysis. 
The pilot interviews enabled us to try a handful of approaches to orient the user's attention to the information they needed to answer their research question. 
This design process resulted in \texttt{MetaExplorer}'s sequential questionnaire format.

A major theme in pilot interview sessions was the difficulty of judging the quality of evidence presented in a study and its applicability to the participant's research question. 
Participants pointed out that these kinds of judgments were typically considered optional, corroborating findings from formative work that meta-analysts tend not to formally assess quality of evidence~\cite{kale2019decision}. 
One participant noted the inadequacy of existing tools for quality assessment: \textit{``these judgments are not black and white'' (Pilot05).}
This inspired us to synthesize existing quality assessment scales (Section 2.1) into a questionnaire allowing ambiguous `not sure' responses.

Sharing intermediate versions of \texttt{MetaExplorer} with experts helped us choose among possible design strategies for handling epistemic uncertainty.
For example, an intermediate version of the tool lacked a triage process for epistemic uncertainty and instead displayed all elicited risks of bias alongside study results in an interactive visualization, explicitly depicting different sources of epistemic uncertainty for the user to explore.
We shared this version of \texttt{MetaExplorer} with two experts, a scientist in the Navy and a professor at a major research university.
Although they agreed that this design made epistemic uncertainty explicit, they also said it undermined confidence in the ability to produce a useful meta-analytic estimate. 
Experts requested a triage process (see Section 4.1.4) where users can express their \textit{``first gut feel'' (Expert01)} about how important potential sources of bias might be and whether they can be resolved before viewing study results.
For this reason, we pivoted our design to focus on helping users reduce epistemic uncertainty, rather than finding more elaborate ways to display it.
\section{System}
We present \texttt{MetaExplorer}, a prototype meta-analysis tool that elicits sources of epistemic uncertainty in literature review and propagates them alongside quantitative study results during meta-analysis.
Readers can find the source materials for \texttt{MetaExplorer} at \url{https://github.com/kalealex/meta-explorer}.

\subsection{Exposition \& use case scenario}
\textit{Scenario:} To demonstrate the \texttt{MetaExplorer} workflow, we return to Kara, the hypothetical scientist investigating the influence of social robots on older adults' mental health (Section 1).

\begin{figure*}
    \centering
    \includegraphics[width=\textwidth]{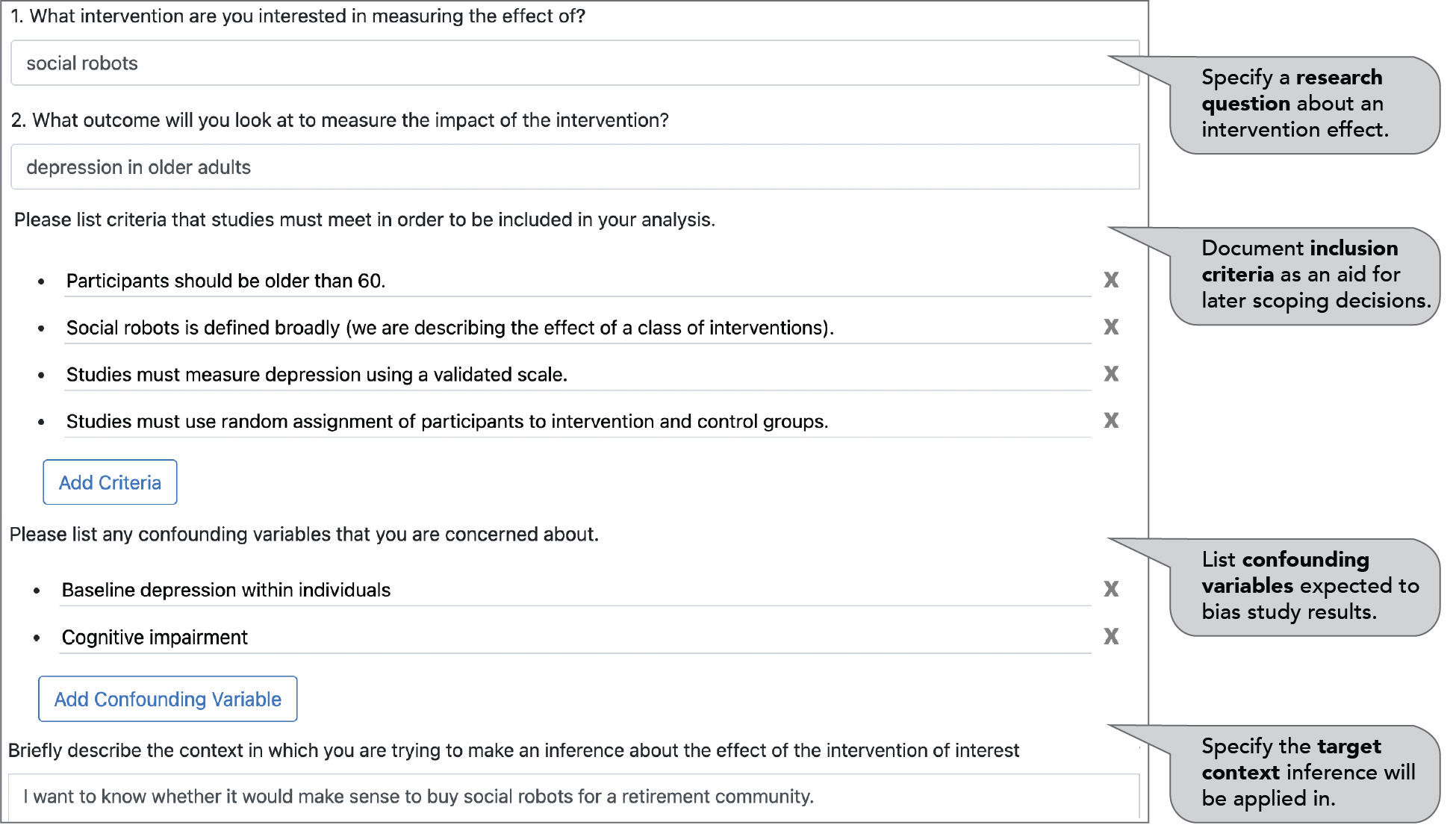}
    \caption{Reviews in \texttt{MetaExplorer} start with the \textbf{Scoping view}, where the user documents choices that will guide what evidence gets included in their summary of the scientific literature.}
    \label{fig:scoping}
    \Description{Scoping a review: MetaExplorer elicits a research question in terms of an intervention and outcome of interest; documenting a user-defined list of inclusion criteria serves as an aid to later decisions about which study results are in scope; documenting a user-defined list of expected confounding variables guides risk of bias judgments during evidence extraction and triage; eliciting the target context as an open text response guides applicability judgements during evidence extraction and triage.}
\end{figure*}

\subsubsection{Scoping}
The \texttt{MetaExplorer} workflow begins with a view where users express research topics and possible research questions per topic.
\texttt{MetaExplorer} supports research questions of the form, ``What is the impact of \texttt{<intervention/>} on \texttt{<outcome/>}?'' which are typical for meta-analyses (\autoref{fig:scoping}, first row).
Next, the user describes what will count as evidence by specifying study inclusion criteria, potential confounding variables, and the target context to which the meta-analytic inference will be applied (\autoref{fig:scoping}, second-fourth rows).
Like pre-registration, answering these questions helps users focus their review and creates a mechanism for personal accountability, something they can check when unsure about how to handle a study.
However, unlike pre-registration, users can return to this page and edit the scope of the review as they review the literature.

\textit{Scenario:}
Kara uses \texttt{MetaExplorer} to document the target context for her inference (a retirement community interested in social robots for mental health; \autoref{fig:scoping}, fourth row), so she can focus on applicable papers. Kara specifies her research question, ``What is the impact of social robots on depression in older adults?'' since depression is the most commonly measured outcome in the sparse literature that quantitatively measures the impact of social robots. Kara scopes her review using inclusion criteria---e.g., documenting that `social robots' refers to a class of interventions rather than a specific robot. She notes concern about study results that fail to control for baseline depression. 

\begin{figure*}
    \centering
    \includegraphics[width=\textwidth]{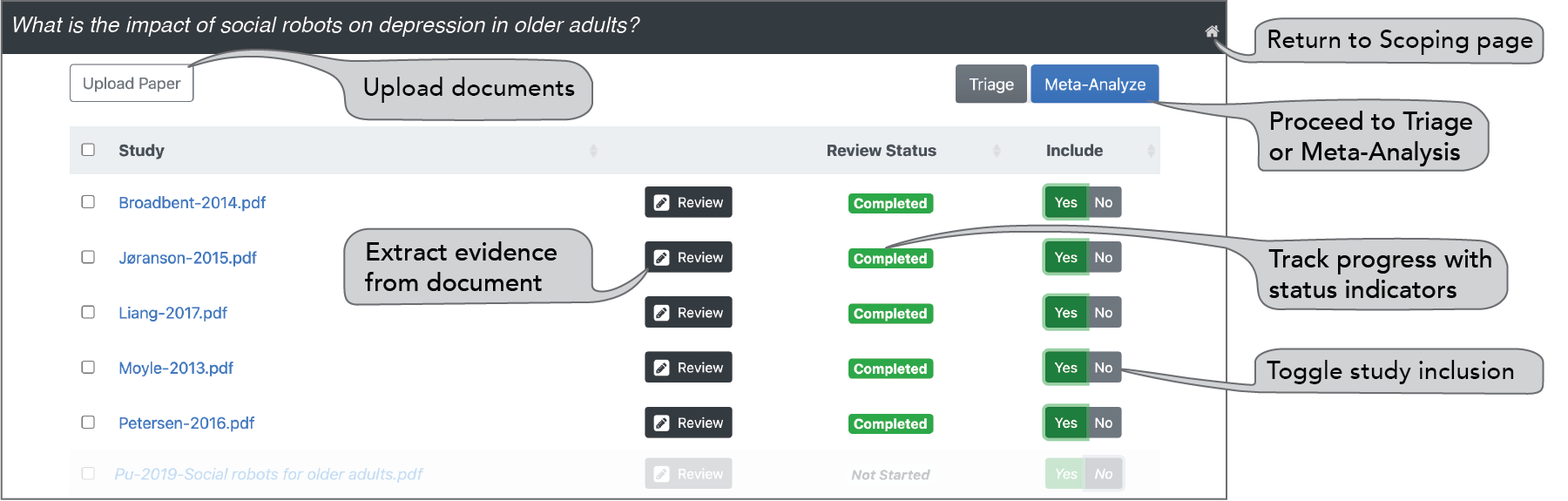}
    \caption{The \textbf{Review management} view in \texttt{MetaExplorer} is a tabular interface where users can upload PDFs to a database, track their progress in reviewing each document, and toggle the inclusion or exclusion of each article.}
    \label{fig:review-management}
    \Description{Review management: reminder of research question at the top of the page; navigation button to return to scoping view; button to upload papers; table with one row per uploaded paper where each row contains a status indicator (complete, not started, etc.), buttons to go to evidence extraction view, or toggle study inclusion in the corpus for meta-analysis; buttons at the bottom of the page to proceed to triage or meta-analysis.}
\end{figure*}

\subsubsection{Review management}
After scoping, users begin to review scientific articles on their topic.
\texttt{MetaExplorer} provides a review management view that enables users to upload articles, toggle provisional study inclusion choices, and navigate between system components for reviewing literature, triaging epistemic uncertainty, and meta-analysis (\autoref{fig:review-management}).

\textit{Scenario:} To save time searching for papers, Kara decides to start by replicating an existing meta-analysis~\cite{Pu2018-robots-meta-analysis}, thus she already has documents to upload. 
For a new meta-analysis, Kara would need to search for articles via online databases and citation networks. Kara uses the review management view to navigate between interfaces for evidence extraction, triage, and meta-analysis.

\begin{figure*}
    \centering
    \includegraphics[width=\textwidth]{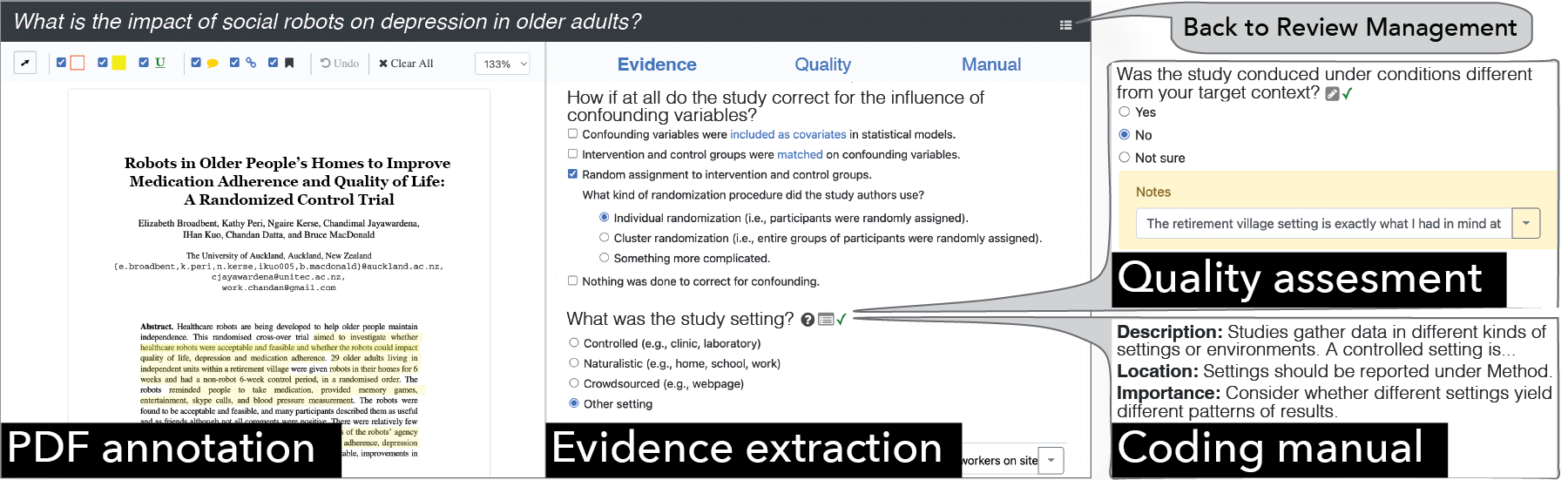}
    \caption{The Evidence extraction tool in \texttt{MetaExplorer} is where users pull effect size information from each study in their review and document concerns about epistemic uncertainty. The major components of this tool are \textbf{PDF annotation} where users read and markup documents, \textbf{Evidence extraction} where \texttt{MetaExplorer} elicits information about the study design and results, \textbf{Quality assessment} where \texttt{MetaExplorer} elicits judgments of epistemic uncertainty, and a \textbf{Coding manual} which guides evidence extraction.}
    \label{fig:evidence-extraction}
    \Description{This overview shows different elements of MetaExplorer’s literature review interface and describes each. The interface contains: a PDF annotation view on the left side with controls for PDF annotation such as drawing rectangles, highlighting, underlining, commenting, linking, bookmarking, undoing, and clearing all annotations; a navigation button to return to review management; and a dynamic web form on the right side with buttons to navigate between different tabs in the form. The evidence extraction tab guides users through the review of each article via and interactive web form. The form branches such that questions about study design at the top of the form tailor questions later on about how results are reported in the document. The evidence extraction form culminates in an evidence table where the user fills in reported effect size information. The quality assessment tab contains questions about epistemic uncertainties that are interleaved with evidence extraction questions through clickable icons that link related questions across these two forms. The coding manual tab contains descriptions for each evidence extraction question in terms of what the user should be looking for, where to find it in the paper, and why it is important in a typical review.}
\end{figure*}

\subsubsection{Evidence extraction}
Literature review happens in \texttt{MetaExplorer}'s evidence extraction tool.
Its major components facilitate (1) annotating documents, (2) recording how a study was run and its results, (3) recording sources of epistemic uncertainty through quality assessment, and (4) checking terminology and coding procedures that come up in scientific review.
The interface is a split view with a PDF annotation tool on the left and a dynamic web form on the right (\autoref{fig:evidence-extraction}) containing three navigation tabs: evidence extraction, quality assessment, and coding manual.

\textbf{The PDF annotation tool} enables users to highlight, draw boxes, underline, and comment on the PDF, and to bookmark and link selected locations in the PDF document (\autoref{fig:evidence-extraction}, left column). We designed this tool after observing how participants used printed articles during pilot interviews.

\textbf{The evidence extraction form} guides the user through coding each article in a meta-analysis (\autoref{fig:evidence-extraction}, middle column). The form includes sections about study identity (i.e., authors, year, title), study context (e.g., what was the study design? What were the mechanisms for experimental control?), participants (i.e., who were the participants? How were they recruited?), measurement (e.g., how were variables defined and measured? What comparisons are reported in the article?), and effect size (i.e., what statistics should be used in a meta-analysis?). 
The form is dynamic: user responses to questions about study design and measurement determine what statistical information the tool asks for. For example, if the user indicates that a study result adjusts for potential confounding variables, the form will later ask which covariates were adjusted for.
Evidence extraction culminates in an evidence table used for data input. The evidence table asks for only the statistics presented in the article which are needed for meta-analysis.

\textbf{The quality assessment form} asks the user to judge the quality of evidence presented in a given article (e.g., \autoref{fig:evidence-extraction}, top of right column) as they move through the evidence extraction form. The form includes sections on risk of selection bias (e.g., did the study fail to control for important confounding variables?), measurement issues (e.g., did the study use a validated measurement scale?), and applicability (e.g., are the participants different from the population the user would like to make an inference about?).
Each quality assessment question is linked to a related question in the evidence extraction form, such that users can navigate quickly between related questions across the two forms.

\textbf{The coding manual} provides explanations of questions in the evidence extraction form for new users (e.g., \autoref{fig:evidence-extraction}, bottom of right column). 
Upon clicking beside a question, users receive a \textit{description} of the question with links to online definitions for necessary jargon; a \textit{location} in the article where the user might find this information; and a brief explanation of the question's \textit{importance} in a typical review.

\begin{figure*}
    \centering
    \includegraphics[width=\textwidth]{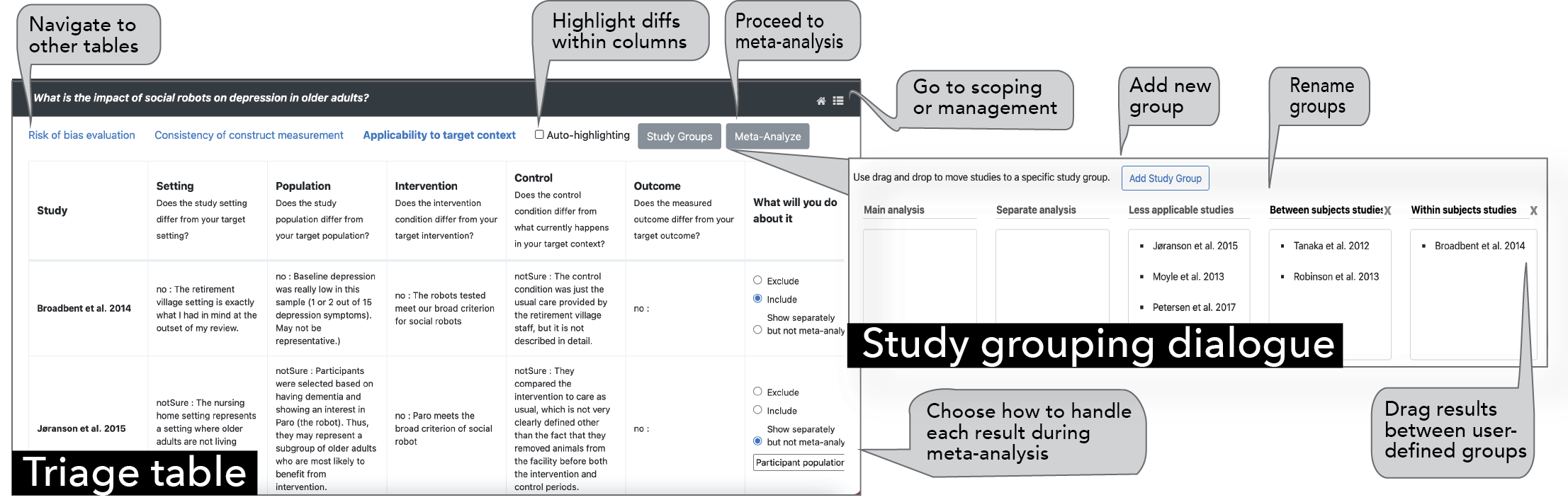}
    \caption{This \textbf{Triage table} presents judgments about the applicability of study results to the target context, which were elicited during quality assessment. \texttt{MetaExplorer}'s triage view also generates similar tables for judging risk of bias and consistency of construct measurement. The \textbf{Study grouping dialogue} gives users control over the grouping of study results in meta-analysis. Actions selected by the user in the rightmost column of the triage table generate default groups (main analysis, separate analysis, and less applicable studies). Study grouping allows users to edit these defaults.}
    \label{fig:triage-applicability}
    \Description{Triage tables: buttons to navigate back to the scoping view or to proceed to meta-analysis; additional buttons to navigate between triage tables for risk of bias, consistency of construct measurement, and applicability; buttons to toggle auto-highlighting of differences within table columns and to access study grouping dialogue. The applicability table is shown for reference: each table row represents a study result, and each column shows the users response to a quality assessment question about applicability of evidence to the target context that the user would like to make an inference about; the rightmost cell in each row contains radio buttons for each study result asking the user to include, exclude, or show [the study result] separately by not meta-analyze. The study grouping dialogue is also shown: an interface containing columns for each user-defined study group (main analysis, separate analysis, and less applicable studies by default); the user can drag tiles representing each study result between these groups; the user can click a button to add new groups, and can rename groups by clicking on the column headers; users can remove groups by clicking an ‘X’ next to the study headers.}
\end{figure*}

\textit{Scenario:} For each article in her review, Kara follows the guided process for evidence extraction. She fills her database with the necessary quantitative information and develops a sense of where the interpretation of the literature is uncertain. For one study, she notes that the authors omitted information about participants, making it difficult to say whether there is selection bias. Another study failed to control for individual differences in baseline depression, a confounder she is concerned about \textit{a priori}. Kara notes that most studies in her corpus recruited participants with dementia, who live less independently and experience greater cognitive decline than the people she intends to make an inference about. She will need to consider these concerns in making a statistical inference. Without \texttt{MetaExplorer}, Kara might have written these concerns down in her lab notebook and forgotten about them, or found them hard to reconcile, upon meta-analyzing her corpus~\cite{kale2019decision}.

\subsubsection{Triage \& study grouping}
\texttt{MetaExplorer} provides a triage process to help users reduce elicited epistemic uncertainty into a set of considerations they believe should guide statistical inference after they complete excluding or reviewing studies in the review management view.
The triage view includes three tables: (1) risk of bias, helping the analyst avoid potentially misleading evidence, (2) consistency of construct measurement, helping the analyst interpret estimates arising from different procedures, and (3) applicability, helping the analyst reason about generalizing study results to their target context. 
Triage tables contain one row per study result and one column for each relevant question from the evidence extraction and quality assessment forms (e.g.,~\autoref{fig:triage-applicability}). 
Each table corresponds to different challenges that come up in meta-analysis (Section 2.2) and each challenge warrants a different \textit{action}.
The rightmost column of each table asks the user, \textit{``What will you do about it?''} with radio buttons that enable the user to include, exclude, or---depending on the triage table---flag results for risk of bias, group results into separate analyses based on what they seem to measure, or show results from less applicable studies without meta-analyzing them.
\texttt{MetaExplorer} auto-highlights differences between cells in each column to draw the user's attention to discrepancies between study designs. 
The triage view also provides a drag-and-drop dialogue (\autoref{fig:triage-applicability}, right) for creating and naming study groups for meta-analysis and dragging results between these groups.
The outputs of the triage process are study groups to be meta-analyzed separately and flags summarizing concerns about potentially biased results.

\textit{Scenario:} Kara uses triage to get an overview of her corpus and decide how evidence should be combined in a meta-analysis. In the risk of bias table, she places flags on two studies which may not have controlled for confounding variables. In the consistency of construct measurement table, she sees that her review contains both within- and between-subjects study designs, which she decides to analyze separately because within-subjects effects represent a different construct (i.e., average treatment effect on an individual) than between-subjects effects (i.e., average treatment effect in a population). In the applicability table, Kara sees that many studies in her corpus recruited only participants with dementia, which is not the population she wants to make an inference about. She decides to view the results of these less applicable studies separately without meta-analyzing them. 
The review that Kara replicated~\cite{Pu2018-robots-meta-analysis} did not separate within- versus between-subjects results, and combined evidence across populations of participants with versus without dementia.
Kara realizes that her meta-analysis will yield a set of contextualized estimates rather than a single estimate that averages over many potentially important variations in the corpus. Although this will make her result less concise, she thinks it is a more realistic summary of available evidence.

\begin{figure*}
    \centering
    \includegraphics[width=\textwidth]{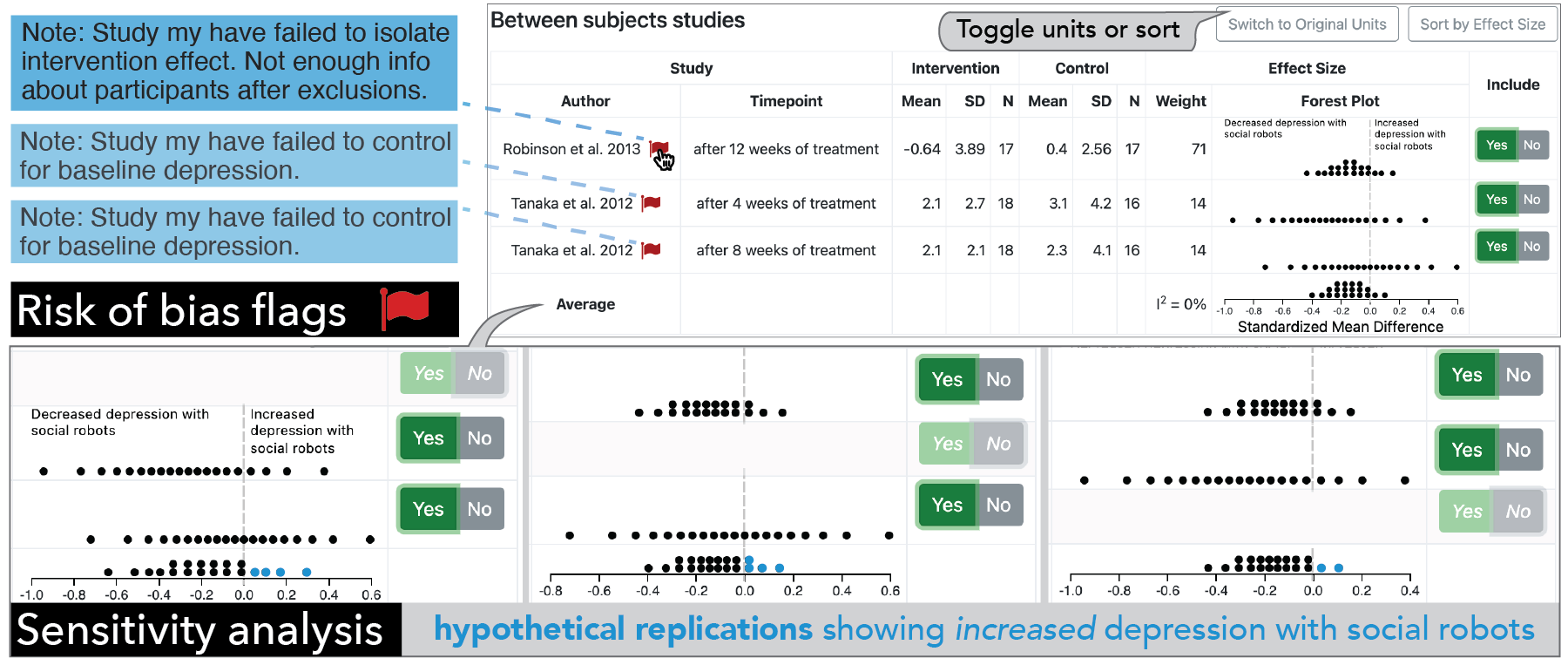}
    \caption{\texttt{MetaExplorer}'s visualization displays summarized epistemic uncertainty alongside quantitative evidence. Mousing over \textbf{risk of bias flags} shows user-generated annotations from triage in a tooltip. Clicking toggle buttons in the table facilitates interactive \textbf{sensitivity analysis} to examine how study inclusion choices impact averages within study groups defined in triage. Quantile dotplots frame each sampling distribution of effect estimates as 20 hypothetical replications drawn from of a given population of studies.}
    \label{fig:forest-plot}
    \Description{Meta-analysis and visualization: each study group from triage gets its own table with one row per study result; table columns give the study’s authors, timepoints of measurement, summary statistics for the intervention and control group, and a quantile dotplot showing the distribution of inferential uncertainty in the study result; the bottom row in the table shows the average effect size in its own quantile dotplot, computed through meta-analysis; controls in the rightmost column enable the user to toggle whether or not each result gets included in the meta-analysis to support sensitivity analysis (e.g., the figure shows pop-outs of three alternative meta-analyses excluding one study at a time, and the results vary such that anywhere from 2 to 4 hypothetical replication studies out of 20 would be expected to find an increase in depression associated with social robots. this shows that the effect of social robots is not robust to study inclusion decisions); where the user flagged study results for risk of bias during triage, flag icons appear in the corresponding row, and the user can read a summary of their concerns in a tooltip by mousing over the flag icon (e.g., 'Note: study may have failed to control for baseline depression.'); buttons at the top of the table enable sorting table rows by the magnitude of the result, and converting results to display in the original units of measurement rather than standardized effect size units which are typically used to compare results across studies in meta-analysis.}
\end{figure*}

\subsubsection{Meta-analysis \& visualization}
The interactive visualization summarizes all results included in a review and enables the analyst to perform sensitivity analysis, assessing how the estimates from meta-analysis change depending on the set of study results included in the model. 
This final step facilitates quick explorations of simple meta-analytic models in light of epistemic uncertainty documented during review.
The D3-generated~\cite{d3js} \texttt{MetaExplorer} visualization is modeled after a forest plot (e.g., \autoref{fig:forest-plot}). 
Each study group defined in triage gets its own table, including the group of less applicable studies that are shown but not meta-analyzed.
Each table row contains summary information about a specific study result alongside a quantile dotplot~\cite{Kay2016} showing the quantitative result in standardized effect size units. At the top of each forest plot are buttons to ``Sort [table rows] by effect size'' and to ``Convert [study results] to original measurement units'', adding independent \textit{x}-axis scales in each row to show non-standardized effects rather than the standardized effect sizes~\cite{Coe2002} typically used in meta-analysis, since non-standardized effect size can provide important context and more robust estimates under certain conditions~\cite{baguley2009standardized}.
Effect sizes supported in \texttt{MetaExplorer} include non-standardized and standardized mean differences for continuous measures as well as risk differences and log odds ratios for dichotomous measures.
Each table uses a common \textit{x}-axis scale to facilitate comparisons across rows of the forest plot, and the bottom row in each shows the meta-analytic average effect size within the study group.
A flag icon appears on rows where the user flagged the results for risk of bias, which users can mouseover to see a description of their concerns.
The rightmost cell in each row contains a toggle button for conducting sensitivity analysis by including or excluding results. 

\textit{Scenario:} \texttt{MetaExplorer}'s visualization shows Kara three forest plots corresponding to her three study groups. One group shows between-subjects results (shown in \autoref{fig:forest-plot}) which were flagged for risk of bias. Although these results might suggest that social robots reduce depression in older adults, Kara explores the space of possible inferences, using sensitivity analysis to determine that the meta-analytic average is not robust to inclusion choices. The results of the within-subjects comparisons, shown in a second forest plot, seem inconclusive. Kara inspects her third forest plot of less applicable studies that were not meta-analyzed. She sees mixed results in studies that recruited participants with dementia. It seems like there is little evidence in this literature that social robots reduce depression in older adults. 
Although this conclusion is in line with the meta-analysis Kara replicated~\cite{Pu2018-robots-meta-analysis}, she can now provide better reasons for her clients about \textit{why} investing in social robots would be premature given current scientific evidence.
The original meta-analysis averaged all of these study results together, suppressing epistemic uncertainty which gives these results meaning to produce a single estimate, whereas Kara's analysis with \texttt{MetaExplorer} produced groups of results informed by epistemic uncertainty that better describe how the literature is methodologically scattered and empirically inconclusive. 
While she is disappointed not to have a single straightforward result to report to her client, Kara thinks that revealing the messiness of the literature is an honest result, which both suggests opportunities to improve future research and answers her practical question. 
\section{Qualitative evaluation of \texttt{MetaExplorer} with meta-analysts}
To evaluate \texttt{MetaExplorer}, we conducted a qualitative user study with 12 scientists knowledgeable about using meta-analysis for different ends across a range of disciplines. 
We structured our evaluation as a guided tour of \texttt{MetaExplorer} during which we interviewed potential users about how they saw the tool supporting their specific meta-analysis workflows.
We synthesized the results of these interviews into a set of themes capturing appraisals of \texttt{MetaExplorer} along multiple dimensions (e.g., usability, trust) as well as remaining challenges and opportunities in designing for meta-analysis.

\subsection{Participants}
We recruited 12 knowledgeable meta-analysts, without overlap with previous participants from the design process (Section 3.1), for our interviews. 
This was a convenience sample recruited from our professional network via email and Twitter.
All participants had sufficient previous experience conducting meta-analyses to inform workflow preferences and other valuable perspectives about scientific review.
Participants were academic researchers in eight countries, mostly in Europe and North America.
Three participants study technology, four study education, two study biological science, one is a cognitive scientist, and two are quantitative methodologists.
The sample composition reflects the intended users of \texttt{MetaExplorer}, scientists across a variety of domains with previous experience conducting meta-analysis.

\subsection{Interviews}
The interviews were structured as a guided tour of \texttt{MetaExplorer}. 
We held interviews on Zoom and saved recordings of each interview for subsequent analysis.
In the first 40-50 minutes of each interview, the interviewer walked participants through the functionality and workflow of \texttt{MetaExplorer} in detail to get their feedback about workflow and features.
We instructed participants to, \textit{``Please speak up if you have impressions about how various software features might be useful to you or how they might create barriers to your work.''}
In the last 10-20 minutes of each interview, the interviewer asked participants two high-level questions to prompt discussion about \texttt{MetaExplorer}.
First, the interviewer asked, \textit{``What merits and drawbacks to you see in a guided process for meta-analysis?''}
Second, the interviewer asked, \textit{``Does \texttt{MetaExplorer} change the way you think about epistemic uncertainty in meta-analysis? If so, in what ways?''}

\subsection{Qualitative analysis}
The first author reviewed and coded video recordings from all 12 interviews.
We adopted a lightweight coding scheme to analyze what participants said about \texttt{MetaExplorer} and meta-analysis more broadly, starting with open codes describing what we discussed with participants.
We used deductive labels for \textit{affordances (A)} and \textit{drawbacks (D)} of \texttt{MetaExplorer}, as well as \textit{feature requests (FR)} and \textit{pain points in current practice (PP)}. 
We also labeled open codes inductively based on the topics that participants frequently raised: \textit{epistemic uncertainty (EU)}, \textit{usability (U)}, \textit{collaboration (C)}, and \textit{domain specificity (DS)}.
For codes associated with a concise and interesting quote, we transcribed the relevant portion of the recording. 
We iteratively grouped open codes and quotes into themes and tensions following an affinity diagramming procedure.

\subsection{Results}
\textbf{Overview of results.} Our qualitative analysis surfaced two primary themes: (1) resolving versus propagating epistemic uncertainty, and (2) imposing structure on scientific workflows for which no normative process is available. 
Participants' comments generally supported our design hypothesis that providing a guided process to elicit and resolve human judgments of epistemic uncertainty should contribute to more trustworthy meta-analyses.
Participants envisioned the role of \texttt{MetaExplorer} as building confidence in the review by assisting humans in finding \textit{``what to compare with what and which data to extract'' (P03)}
and \textit{identifying the difficult spots in the coding sheet (P11)}, which reflect epistemic uncertainty that may need to be resolved by a team of coders.
Guiding human judgments and making them explicit in \texttt{MetaExplorer} facilitates open analyses that can be shared with colleagues for audit \textit{(P01, P12)}, which aids in socially distributed construction of knowledge and trust across networks of scientists~\cite{Longino1990}.
However, different standards and procedures are meaningful in different scientific disciplines.
As a consequence, participants disagree about whether \texttt{MetaExplorer} expects a review procedure that is too rigid \textit{(e.g., P02, P08)} or not rigid enough \textit{(e.g., P06)}.
This speaks to challenges and opportunities---both specific to meta-analysis and broadly applicable---in creating software that encourages researchers to adopt new practices as part of statistical reform.

\subsubsection{Resolving versus propagating uncertainty}
Participants often questioned whether to resolve or propagate epistemic uncertainty, by either dismissing sources of uncertainty as negligible or summarizing concerns for further consideration later in analysis.
We summarize observations on when participants face this choice, how \texttt{MetaExplorer} might help, and challenges that make this choice difficult to design for.

\textbf{Shades of gray in scoping decisions.}
Deciding whether to resolve or propagate epistemic uncertainty surfaced primarly as participants considered \texttt{MetaExplorer}'s support for scoping decisions. 
Participants described how they draw boundaries around their corpus to meta-analyze enough evidence to be informative without including so much variety as to obfuscate their inference.
This balance becomes difficult when the literature is sparse or heterogeneous in construct definitions or measurements \textit{(P01)}, which only becomes clear \textit{``once you've looked at a dozen on more studies'' (P12)}.
Participants comments supported our hypothesis that the scope of a meta-analysis cannot be fixed or `preregistered' from the outset of a review, but instead must be reconsidered throughout the review.

Multiple participants commented that \texttt{MetaExplorer} helps to track the evolving scope of a review.
One participant \textit{(P03)} typically keeps a notebook of scoping decisions and remarked that \texttt{MetaExplorer}'s scoping page would be more systematic.
Another participant \textit{(P05)} bemoaned how ad hoc workflows for handling epistemic uncertainty can erode sense of scope and introduce mission drift about the goals of meta-analysis.
Without a tool like \texttt{MetaExplorer} to document scope and guide study inclusion decisions, participants \textit{(e.g., P04, P05)} must invent their own systems of organization and accountability. 

Participants saw \texttt{MetaExplorer} assisting with scoping decisions most directly by asking users about shades of gray in study inclusion---i.e., reasoning about \textit{``if the causal inference [supported by a result] is strong or not.'' (P08)}.
Participants clarified that they typically resolve concerns about applicability by narrowing scope rather than by considering what can be learned from different groups of studies. 
\begin{quote}
    \textit{We don't look at interventions, for example, in students with disabilities if our population of interest is English language learners... The target context becomes part of the inclusion criteria. (P10).}
\end{quote}
However, what it means to generalize can become ambiguous. 
One participant described how in academic research \textit{``There's not always an applied target context.'' (P03)}.
Participants \textit{(P05, P10)} told us that decisions about how to parse the literature are informed by norms, which can feel arbitrary.
\texttt{MetaExplorer} makes these considerations explicit.
\begin{quote}
    \textit{`Does this study fit the context I want to generalize about?' It's something that I've vaguely heard people think about, but it's not something that I've seen anybody put into a tool like this. I think that's great because a lot of meta-analyses are: find everything you can, throw it into a big pot, and stir, and out comes something that is of dubious usefulness for particular purposes, like when you are trying to make decisions. (P12).}
\end{quote}

However, sometimes participants reported there is no satisfying way to scope a review.
For example,
\textit{``We actually shelved this meta-analysis on data literacy tools because... the way that people operationalize data literacy is so varied and diverse that it actually doesn't make sense to compare.'' (P09)}.
With a larger scale of about 60-100 studies, several participants \textit{(e.g., P04)} said grouping results for meta-analysis becomes more difficult, even with \texttt{MetaExplorer}'s triage process, because the number of possible groupings grows with the number of results.

\textbf{Preference for statistical approaches to uncertainty.}
Participants reported preferring to use statistical tools to resolve questions about how and whether measurements should be combined.
For example, some participants valued quantitative feedback for inclusion decisions: 
\textit{``I like having the ability to run sensitivity analysis. Like, if something looks off, how much does it change the results?'' (P08)}.
Many participants \textit{(e.g., P03, P12)} wanted to use hierarchical models to account for how sources of variation are clustered depending on study designs.
By default \texttt{MetaExplorer} applies a separate hierarchical model to each user-defined study group, however, it does not handle special cases where measurements are inherently correlated---e.g., when combining multiple measurements of the same sample.
Because \texttt{MetaExplorer} doesn't enable such complexity in modeling, one participant \textit{(P03)} worried it may not encourage users to be ambitious enough about incorporating a wide variety of evidence into meta-analysis.

Participants disagreed about adding more complex modeling features, but some wanted the reassurance of verifying what models \texttt{MetaExplorer} runs.
\textit{``It may not estimate or run the models the way that I would need to to publish my papers, but I'm not totally sure.'' (P07)}.
Some participants \textit{(P03, P06, P11)} wanted to manipulate the underlying R code. 
In contrast, one methodologist and tool builder \textit{(P12)} recommended not revealing model specifications, acknowledging that this would be inaccessible to less experienced users.
Future tools like \texttt{MetaExplorer} could strike a better balance in model exposure by having an optional view that makes code available for more expert users. 
However, we question whether novice users should rely on meta-analytic models without understanding them.

\textbf{Collaboration.}
Often sources of epistemic uncertainty cannot be resolved through statistical approaches---e.g., when methodological variations do not form clear clusters---and deliberations among colleagues play a crucial role in deciding how to handle a concern.
Participants viewed the tool as a skeptical collaborator in such deliberations.
\begin{quote}
    \textit{I would model this tool to be a grouchy reviewer that constantly convincing me not to publish the study because I don't have a corpus that is good enough, or I don't have enough certainty. (P09).}
\end{quote}
This participant valued \texttt{MetaExplorer} as a way of organizing knowledge to promote reflection.
Another participant expanded on this, remarking that the tool pushes users to discuss what would count as a generalizable inference in the target context.
\begin{quote}
    \textit{Now that I've seen this, I really think that needs to be an integral part of a meta-analysis. I have to admit that in meta-analyses I've been involved in, these conversations didn't come up that much. I don't remember having deep, long conversations about how studies contribute to making policy decisions for particular situations in a particular context. (P12).}
\end{quote}

Participants frequently commented that \texttt{MetaExplorer} would make an excellent collaboration platform. 
Three participants \textit{(P05, P06, P07)} bemoaned the difficulty of finding free literature review tools that support synchronous collaboration.
One participant described how collaborating through reference managers can lead to epistemic uncertainty getting lost in communication.
\begin{quote}
    \textit{Mendeley did a whole lot of heavy lifting for one of the meta-analyses I completed years ago. We just couldn't find anything... It just wasn't streamlined, and it would get really frustrating because inevitably someone would say, `Oh, I left you a note about that three months ago.' (P05).}
\end{quote}
\texttt{MetaExplorer} facilitates progress tracking through indicators in the review management view of what has (not) been coded. 
We envision extending this interface to include action items for collaborators, e.g., assigning people to documents, requesting clarification on codes, or resolving disagreements through \textit{dual review}---independent coding by multiple scientists, which was the most common feature request.

Multiple participants \textit{(P04, P07, P11)} remarked on the affordances of \texttt{MetaExplorer}'s guided process for training teams of coders with mixed levels of experience, and they said that this sort of coordination usually takes a lot of time and energy.
We observe that much of what teams need to train and coordinate about involves the handling epistemic uncertainty (e.g., what needs to be coded to differentiate study groups).
\texttt{MetaExplorer} provides workflows dedicated to handling these concerns and in doing so makes it less likely teams of coders will lose important contextualizing information.

\subsubsection{Imposing structure without a clear normative procedure}
Our analysis of interviews suggests that the primary tension around designing for meta-analysis is how much structure to impose on the process. 
We find a striking contrast between consensus around the need for standardization in research synthesis and participants' idiosyncratic preferences about what standards are meaningful in their domain.

\textbf{Need for structure.}
All participants highlighted the benefits of \texttt{MetaExplorer}'s streamlined process.
Scaffolding document analysis helps users think through coding decisions \textit{(P07)}, structures resulting knowledge \textit{(P09)}, prevents decision paralysis regarding \textit{``what to worry about'' (P10)}, and could reduce variance in results across research teams \textit{(P11)}.

Participants contrasted \texttt{MetaExplorer}'s guided process with their typical, more ad hoc approach.
\textit{``I typically think about [epistemic uncertainty] more manually, less systematically. It comes up all the time, but the tool allows you to have a very strict, very formal way of dealing with it.'' (P02)}.
Another participant echoed,
\textit{``It helps to find weaknesses or blind spots that you hadn't thought about, moreso than if you were to do it more chaotically.'' (P03)}.
Participants \textit{(P04, P11)} mentioned often adding risk of bias items to coding spreadsheets midway through a review, and then re-coding articles for previously \textit{``hidden''} information.
Beyond structuring their thinking, multiple participants \textit{(P05, P08)} appreciated how \texttt{MetaExplorer} backed their work with a relational database, which reduces the time required for data cleaning in meta-analysis, e.g., from months to minutes.
\texttt{MetaExplorer} generates triage tables from this relational database, another data management automation that one participant particularly appreciated.
\textit{``When I was describing the spreadsheet we did, it looked pretty much like this. The fact that this spreadsheet gets generated as I'm doing each review---it's very helpful not to have to do this by hand.'' (P09)}.

\textbf{Need for customization.}
Scientific fields have different ways of designing and reporting studies, so participants frequently requested to tailor \texttt{MetaExplorer} to their domain, similar to customizing codebooks in spreadsheets.
For example, 
\textit{``Would you make this more flexible for people who are in engineering or ecology and evolution? Because our experiments or studies are very much different than social psychology, like a lot of ecology and evolution is observational.'' (P05)}.
This echos concerns from other participants, e.g., that \texttt{MetaExplorer} is geared toward a \textit{``specific type of research design'' (P09)} in ways that rule out qualitative evidence, and that \texttt{MetaExplorer} doesn't support certain standards like PRISMA~\cite{PRISMA}, MUTOS~\cite{becker2017improving,Manski2019-MA}, or PICOTS~\cite{PICOTS} \textit{(P06, P08)}, which are popular in medicine.
On the other hand, some participants \textit{(e.g., P08, P11)} found \texttt{MetaExplorer} sufficiently aligned with the spirit of these standards in encouraging documentation of and reflection about review scope.

One form of document analysis that requires considerable codebook customization is qualitative evidence synthesis.
A common grievance with \texttt{MetaExplorer} \textit{(P04, P05, P10)} was prioritizing quantitative meta-analysis over qualitative systematic review, especially eliciting research questions in terms of causal effects of interventions.
One participant envisioned how \texttt{MetaExplorer} could be extended to support evidence from mixed methods:
\begin{quote}
    \textit{Is there some sort of mapping that I could have between this [forest] plot and the qualitative description of results? What that would show me is why---because here I can see with the forest plot some effect sizes, but I don't know why I am seeing those. If I could click to say, `Show me the thematic analysis for people who were in this group,' or `What was any sort of summary of qualitative coding of interviews with people in this group?' That's something I've never seen. (P09).}
\end{quote}
We discuss ways to realize this vision in Section 6.2. 

One proposed consequence of \texttt{MetaExplorer} not offering enough customizability is that users won't adopt a tool that doesn't cover the same use cases as their hand-rolled procedures, however ad hoc they are \textit{(P04)}.
In developing \texttt{MetaExplorer}, implementing a streamlined process required opinionated choices about supported procedures.
However, the preference among participants to work in spreadsheets despite their problems implies that users will incur substantial time and labor costs to maintain entrenched workflows and practices. 
Interoperability with Excel and more support for on-the-fly procedural modifications might promote widespread adoption of tools like \texttt{MetaExplorer}.

\textbf{Structure as a representation of mental models.}
We interpret the lack of agreement among participants about standards as evidence that scientists' mental models of research synthesis are highly divergent.
\texttt{MetaExplorer} was hit-and-miss in matching these mental models.
For some participants, \texttt{MetaExplorer}'s evidence extraction process seemed to mirror their preferred perspective---e.g., \textit{``This is how it looks in my brain.'' (P05)}.
For other participants \textit{(e.g., P05, P06, P08, P09)}, the guided tour of \texttt{MetaExplorer} elicited requests for different standards.
At the same time, some comments we observed imply that mismatch is often an opportunity.
One participant described how many benefits of \texttt{MetaExplorer} come from users updating their mental models to match the tool.
\begin{quote}
    \textit{A guided tool like this imposes that structure which maybe a person doing a meta-analysis is not thinking about it this way. Maybe they have a different structure in their head which can lead to some discrepancy or tension. But having a tool like this imposes a structure that can be very useful to people doing a meta-analysis if they have not fully set up a structure themselves or just have a vague notion. (P12).}
\end{quote}
Our results point overall to the need to provide users with ways to express their mental models so that tools like \texttt{MetaExplorer} can update reciprocally.
This fits with participants' \textit{(e.g., P09, P12)} conceptualization of \texttt{MetaExplorer} as a partner in collaboration.
\section{Discussion}
Through developing and evaluating \texttt{MetaExplorer} with experienced meta-analysts and meta-analysis experts (25 people total), our work advances the design of software for promoting awareness of sources of epistemic uncertainty in meta-analysis that get dredged up during literature review but are seldom propagated to resulting inferences~\cite{kale2019decision}.
In particular, \texttt{MetaExplorer}'s features for structuring scoping and triage decisions and conducting sensitivity analysis through interactive forest plots were successful design strategies.
Our interviews with potential users suggest that \texttt{MetaExplorer}'s emphasis on epistemic uncertainty might result in meta-analyses that better characterize heterogeneity in scientific literature, rather than averaging over disparate results.

Our research on \texttt{MetaExplorer} also points to design implications beyond meta-analysis. 
Data analysis tools writ broadly might benefit users by 
\textbf{guiding documentation of and direct consideration of how to address epistemic uncertainty}, e.g., by systematically resolving or propagating descriptive concerns about data quality or meaning alongside statistical information.
Similarly, other interactive systems for data analysis should provide \textbf{data management and workflow automations}, since participants claim these are instrumental for accelerating and systematizing documentation and triage of data quality concerns.
However, these automations must be configurable (see Section 6.2).
A particularly important lesson from designing \texttt{MetaExplorer} is that \textbf{epistemic uncertainty should be summarized according to a predefined workflow rather than explored in an open ended fashion} because (1) scientists tend to have principled ways of handling specific sources of epistemic uncertainty \textit{a priori} and (2) open exploration of epistemic uncertainty can promote a form of decision paralysis where the breadth of reasonable interpretations of data is exaggerated by a non-reduced overview. 
In particular, we expect these principles to be useful for data analysis and communication settings that involve aggregating evidence under hard-to-quantify epistemic uncertainty, such as forecasting applications (e.g.,~\cite{gelman2020information,padilla2021uncertain}) or combining different forms of evaluative information to assess models (e.g.,~\cite{hutchinson2022evaluation}).

\subsection{Limitations}
Developing a sufficiently flexible document analysis interface with appropriate scope for a prototype required us to make opinionated decisions, such as tailoring \texttt{MetaExplorer} to handle controlled experiments rather than a wider variety of study designs. 
While necessary, these scoping decisions limit \texttt{MetaExplorer} to reviews that terminate in meta-analysis, which is not appropriate when available evidence does not support causal inferences or the user wants another form of evidence summary. 

Additionally, a more formal evaluation, where users conduct a meta-analysis with \texttt{MetaExplorer} and the quality of their inferences is assessed rigorously, would allow us to say whether design patterns in \texttt{MetaExplorer} will improve the quality of inferences in actual use compared to current practices. 
However, our experiences suggest this may be hard to realize in practice due to (1) the difficulty of benchmarking user performance when the core tasks in \texttt{MetaExplorer} involve seemingly ``subjective'' contextually-dependent judgments, and (2) the challenges of recruiting meta-analysis experts to use a tool for the extended time period that meta-analysis tends to require. 
We opt for guided tour interviews because we seek \textit{holistic} feedback on \texttt{MetaExplorer}, and given the many tasks involved in scientific review and meta-analysis, other methods we considered (e.g., think-aloud, case studies) would have required more time than our participants could offer.

\subsection{Future work}
Our interviews surfaced opportunities for future work extending a system like \texttt{MetaExplorer} for \textbf{supporting collaborative document analysis}, such as by adding functionality for assigning individual users to review specific documents and resolving disagreements between reviewers.
Review assignment could be handled in \texttt{MetaExplorer}'s review management view using a tagging system to request an individual's attention on a document, and using personalized progress indicators and to-do lists to guide each user's attention. 
Disagreements between reviewers could be resolved in a tabular interface similar to the triage view, generated automatically from a database but customizable to subsets of questions, that would show disagreements across independent reviews of the same document. 
These refinements would make the social aspects of analysis decisions explicit in \texttt{MetaExplorer}, enabling users to calculate disagreement statistics and to better resolve ambiguity about the specific statistics they need to extract for meta-analysis.

Future work might also \textbf{add qualitative results to \texttt{MetaExplorer}} (e.g., thematic analysis), 
relaxing the assumptions that \texttt{MetaExplorer} makes about what should be considered evidence and giving users more flexibility to define appropriate standards for their review. 
Doing so requires changes to the evidence extraction interface, the triage tables, and the forest plot. 
During evidence extraction, users need a way to select which questions are mandatory to answer in order for the review to be marked complete. 
Users also need to be able to add custom questions that are tailored for the specific research design of the study. 
During triage, users need ways to search, sort, and filter study results in order to more easily cluster studies according to what and how they measure.
Participants also suggested an overview visualization of current study groupings and the ability to add or remove columns from the default 
triage table layout.
These changes would make it more feasible to organize qualitative evidence across studies, and could also improve the triage process for quantitative evidence at the scale of 100 studies.

Summarizing qualitative evidence would also require incorporating additional contextualizing information into the \texttt{MetaExplorer} visualization.
For example, we might add word clouds or additional annotations to summarize coding schemes from qualitative analysis, perhaps highlighting common codes or themes across analyses.
This would improve the affordances of the \texttt{MetaExplorer} visualization for propagating sources of epistemic uncertainty, providing a more flexible mapping between sampling distributions and qualitative claims.

\textbf{Progressive form customization}, or just-in-time form branching, is a promising way to support greater flexibility in the evidence extraction and quality assessment forms.
\texttt{MetaExplorer} already does some of this---e.g., using questions about study design to filter subsequent questions about what was reported, which in turn determine the layout of the evidence table. 
With \texttt{MetaExplorer}, we demonstrate how this design pattern can be used to cover substantial methodological variation within interventional experiments. 
However, we could extend this design pattern by using \textit{templates} to capture important considerations and contingencies under different kinds of research designs.
These templates would represent questions and contingencies among them, which users could select from on the fly during evidence extraction. 
We could also expose the template formalism to users through an editing interface, enabling them to author templates by composing new questions or combining questions from existing templates.
This would support different standards in quality assessment---e.g., users could opt for the Cochrane Risk of Bias Assessment~\cite{Cochrane-riskofbias} if this is meaningful to their research community.
It would also more formally demarcate the roles of expert users (authoring templates) and novice users (following templates) in collaboration, a use case that was not a design goal for \texttt{MetaExplorer} but which participants described as a pain point in current practice.

Finally, a handful of participants \textit{(P06, P07, P08, P11)} requested ways to \textbf{flexibly specify models} of their choice within \texttt{MetaExplorer}. 
Adding a model editing dialogue to the \texttt{MetaExplorer} visualization where users could modify the default meta-analytic model in a code block would support this.

\section{Conclusion}
We present \texttt{MetaExplorer}, a software prototype for representing and reasoning with uncertainty about scientific literature during meta-analysis.
\texttt{MetaExplorer} is a proof-of-concept exploring new design patterns for propagating unquantified epistemic uncertainty in end-to-end quantitative data analysis.
By prototyping these design patterns in \texttt{MetaExplorer} and eliciting feedback from knowledgeable meta-analysts, we find challenges and opportunities around (1) supporting the documentation, collaboration, and modeling efforts required to resolve sources of epistemic uncertainty and (2) developing widely-applicable yet sufficiently flexible standards around data quality.
\texttt{MetaExplorer} opens the door to new ways to design data analysis software with an emphasis on how unquantified uncertainty informs statistical inference.

\begin{acks}
This work was funded by US Navy STTR Contract N6833517-C-0410 in partnership with Stottler Henke Assoc, and NSF award \#1749266.
\end{acks}

\bibliographystyle{ACM-Reference-Format}










\end{document}